\documentclass[aps,prd, reprint, superscriptaddress,nofootinbib]{revtex4-1}%
\usepackage{amssymb}
\usepackage{amsfonts}
\usepackage{amsmath}
\usepackage{amsthm}
\usepackage{graphicx}
\usepackage{verbatim}
\usepackage[usenames]{color}%

\providecommand{\U}[1]{\protect\rule{.1in}{.1in}}
\definecolor{blue}{rgb}{0,0,1}

\definecolor{red}{rgb}{1,0,0}

\def \a {\alpha}
\def \b {\beta}

\def \d {\delta}

\def \k {\kappa}

\def \m {\mu}
\def \n {\nu}

\def \p {\phi}

\def \th {\hat{t}}
\def \ph {\hat{\phi}}

\begin{document}

\title{Stationary cylindrically symmetric spacetimes with a massless scalar field and a non-positive cosmological constant}
\author{Cristi\'an Erices}
\email{erices@cecs.cl}
\affiliation{Centro de Estudios Cient\'{\i}ficos (CECs), Av. Arturo Prat 514, Valdivia, Chile.}
\affiliation{Departamento de F\'isica, Universidad de Concepci\'on, Casilla 160-C, Concepci\'on, Chile.}
\author{Cristi\'an Mart\'{\i}nez}
\email{martinez@cecs.cl}
\affiliation{Centro de Estudios Cient\'{\i}ficos (CECs), Av. Arturo Prat 514, Valdivia, Chile.}


\begin{abstract}
The general stationary cylindrically symmetric solution of Einstein-massless scalar field system with a non-positive cosmological constant is presented. It is shown that the general solution is characterized by four integration constants. Two of these essential parameters have a local meaning and characterize the gravitational field strength. The other two have a topological origin, as they define an improper coordinate transformation that provides the stationary solution from the static one. The Petrov scheme is considered to explore the effects of the scalar field on the algebraic classification of the solutions. In general, these spacetimes are of type I. However, the presence of the scalar field allows us to find a non-vacuum type O solution and a wider family of type D spacetimes, in comparison with the vacuum case. The mass and angular momentum of the solution are computed using the Regge-Teitelboim method in the case of a negative cosmological constant. In absence of a cosmological constant, the curvature singularities in the vacuum solutions can be removed by including a phantom scalar field, yielding non-trivial locally homogeneous spacetimes. These spacetimes are of particular interest, as they have all their curvature invariants constant.
\end{abstract}

\maketitle

\section{Introduction}

In vacuum, the static cylindrically symmetric spacetimes, in absence of a cosmological constant, was found by Levi-Civita \cite{LeviCivita} just few years after the emerging of General Relativity. However, the inclusion of a nonzero cosmological constant was only achieved almost 70 years later by Linet \cite{Linet} and Tian \cite{Tian}. More recently, some geometrical properties of these spacetimes, such as the presence of conical singularities,  were reviewed in \cite{daSilvaC,Bicak}. The stationary cylindrically symmetric vacuum solution was discovered independently  by Lanczos \cite{Lanczos} and Lewis \cite{Lewis}. The general solution contains a number of integration constant, whose physical interpretation has been studied in \cite{daSilvaA,daSilvaB}. In vacuum,  the cylindrical stationary  spacetime with a nonvanishing  cosmological constant was derived in \cite{Santos} (see also \cite{Krasinski}).  The interpretation of the integration constants was clarified in \cite{MacCallumSantos}, where was proved that three of them are indeed essential parameters. Two integration constant have a topological origin \cite{MacCallum}, and a third one  characterizes the local gravitational field.    

Despite the static cylindrically symmetric spacetimes are widely known in vacuum, exact solutions containing a massless scalar field as matter source have not been obtained in the most general form until now. Only solutions with plane symmetry, which are a particular case of the cylindrical ones,  have been reported \cite{Vuille, GM}.

In this article, the general stationary cylindrically symmetric solution of Einstein-massless scalar field system with a non-positive cosmological constant $\Lambda$ is found, and its geometrical properties are studied. The aim of this work is to determine the implications of a massless scalar field in a cylindrically symmetric system. Due to the high interest in exact solutions whose asymptotic behavior approaches the anti-de Sitter spacetime, we include in the analysis a negative cosmological constant. In fact, the solutions presented here, for $\Lambda <0$, have that asymptotic behavior. Moreover, we study the effect of a massless scalar field in the case of a vanishing cosmological constant, i. e., we explore the backreaction  generated by the scalar field in the well-known Lanczos-Lewis and Levi-Civita spacetimes. 

As is expected, in absence of  suitable potentials and non-minimally couplings for the scalar field, the no-hair theorem rules out solutions having event horizons, and this is precisely our case. We are just considering a massless scalar field with a constant potential (zero or negative). Thus, in general, the solutions  presented here contain naked singularities, which however could have some physical interest \cite{Gubser}.

The article is organized as follows. In the next section,  the field equations are solved by considering a general stationary cylindrically symmetric ansatz and a negative cosmological constant.  We obtain the general solution, which can be expressed as a linear combination of three functions. Then, the local properties of the solutions are studied using the Newman-Penrose (NP) formalism, where the Weyl-NP scalars allow to obtain the Petrov classification of these spacetimes.  It is shown that a parameter included through the scalar field enlarges the family of spacetimes with respect to the vacuum ones. Afterwards, following \cite{MacCallum}, the stationary spacetime is obtained from the static one by means of a topological construction. These formalisms allow us to identify the four essential parameters of the general solution. One of them is the amplitude of the scalar field, which in conjunction with a second one describe the strength of the gravitational field. The remaining parameters have a topological origin and are just globally defined, because they cannot be removed by a proper coordinate transformation.  Moreover, the mass and angular momentum are computed by using the Regge-Teitelboim method \cite{ReggeTeitelboim}. These conserved charges illustrate the physical meaning of the essential parameters. The case of a vanishing cosmological constant is considered in section \ref{four}. We note that it is necessary to integrate the field equations from scratch, because a special class of solutions is not available by just taking the limit  $\Lambda\rightarrow0$ in the solutions presented in  Sec. \ref{two}.  We found that these spacetimes have all their scalar invariants constant, and are supported by a phantom scalar field. The last section contains some concluding remarks.

\section{Stationary cylindrically symmetric solutions with $\Lambda<0$}\label{two}

We consider the Einstein-Hilbert action with a massless scalar field and a cosmological constant  $\Lambda$,
\begin{equation} \label{action}
I=\int d^{4}x\sqrt{-g}\left[\frac{R-2\Lambda}{2 \kappa}-\frac{1}{2}g^{\m\n}\partial_{\m}\Phi\partial_{\n}\Phi\right],
\end{equation}
where $\kappa=8 \pi G$ is the gravitational constant. The stress-energy tensor turns out to be
\begin{equation}
T_{\m\n}=\partial_{\m}\Phi\partial_{\n}\Phi-\frac{1}{2}g_{\m\n}g^{\a\b}\partial_{\a}\Phi\partial_{\b}\Phi,
\end{equation}
and the field equations are given by
\begin{equation} \label{FE}
G_{\m\n}+\Lambda g_{\m\n}=\kappa T_{\m\n},   \quad
\Box \Phi=0.
\end{equation}

The general stationary, cylindrically symmetric\footnote{In order to include spacetimes lacking of a regular axis, we are adopting the less restrictive definition of cylindrical symmetry given in \cite{MacCallumSantos}.} configuration can be described by the line element
\begin{equation}\label{ds2}
ds^2\!=\!g_{tt}(r)dt^2+g_{\p\p}(r)d\phi^2+g_{zz}(r)dz^2+2g_{t\phi}(r)dt d\phi+dr^2,
\end{equation}
where the coordinates range as $t\in(-\infty,\infty)$, $r\in[0,\infty)$, $z\in(-\infty,\infty)$ and $\p\in[0,2\pi)$, and a scalar field depending just on the radial coordinate, $\Phi=\Phi(r)$.  

The general solution \eqref{ds2} of the field equations \eqref{FE}, considering a negative cosmological constant $\Lambda=-3 l^{-2}$, can be written as a linear combination of the functions
\begin{equation} \label{gs}
g_i(r)=\left(\frac{ e^{3 r/l}-b}{ e^{3 r/l}+b}\right)^{K_i}\!\left(e^{3 r/l}
-b^2 e^{-3 r/l}\right)^{2/3}\!,
i=\{0,1,2\},
\end{equation}
where the metric coefficients read
\begin{equation}\label{ds2comp}
\begin{split}
g_{tt}(r)&=a_1 g_1(r)-a_0 g_0(r),\\
g_{\p\p}(r)&=b_1 g_1(r)-b_0 g_0(r),\\
g_{t\p}(r)&=\sqrt{a_0 b_0}g_0(r)-\sqrt{a_1 b_1}g_1(r),\\
g_{zz}(r)&=c_0 g_2(r).
\end{split}
\end{equation}
The scalar field is given by
\begin{equation}\label{sf}
\Phi(r)=\Phi_0+\frac{1}{2}\sqrt{\frac{\a}{2\k }}\log\left(\frac{ e^{3 r/l}-b}{ e^{3 r/l}+b}\right)^2.
\end{equation}
Here $K_i$, $a_0$, $a_1$, $b$, $b_0$, $b_1$, $c_0$, $\alpha$ and $\Phi_0$ are integration constants. The constants $K_i$ are not independent, since they verify the algebraic relations
\begin{eqnarray}
K_0+ K_1+ K_2&=&0,\label{relnorm}\\
K_0 K_1+ K_1 K_2+ K_2 K_0&=&-\frac{4}{3}+\a.\label{relesp}
\end{eqnarray}
In order to ensure a real metric and scalar field, the previous algebraic relations fix bounds for the constants. The constant $\alpha$ runs in the interval $0\leq\alpha\leq 4/3$,  and the constants $|K_i|$ are bounded from above by  $\frac {2} {3}\sqrt {4  - 3\alpha}$, $\frac {1} {3}\sqrt {4 - 3\alpha}$, and $\frac {1} {3}\sqrt {4 - 3\alpha}$ in any order. 

Note that the presence of the scalar field is encoded in the additional integration constant  $\alpha$ in \eqref{relesp}.  In absence of the scalar field, the stationary solutions presented in \cite{Santos},  and the static ones in \cite{Linet,Tian}, are recovered.

The constant $c_0$ can be absorbed by rescaling the noncompact coordinate $z$, and only one of the constants $a_0$, $a_1$, $b_0$, $b_1$ is essential, as it will become clear in the next subsection.  In order to get insight about the parameter $b$, it is convenient to start with static metric
\begin{equation}\label{Linet1}
ds^2=-g_{0}(r)dt^2+g_{1}(r)l^2 d\phi^2+g_{2}(r)dz^2+dr^2.
\end{equation}
The constant $b$ determines the location of the axis of symmetry at $r_0=l/3\log |b|$, and it can be removed from the scalar field by a shift of the radial coordinate $r \rightarrow r+r_0$. With this shift,  $b$ just appears as a multiplicative factor $b^{2/3}$ in $g_{i}$,  and consequently, the invariants do not depend on $b$.  In other words,  $b$  could be removed from the solution by rescaling the coordinates $t, z, \phi$. However, $\phi$ is a compact coordinate and global properties will be modified with this rescaling. In fact,  the metric with the shifted radial coordinate reduces in absence of the scalar field to that shown in \cite{Bicak}, where a conicity parameter equivalent to $b^{-1/3}$  is explicitly exhibited. In summary, $b$ has no relevance for the local properties, but it is a topological parameter that contributes to the mass of the solution (see subsection \ref{three}). 

The general solution previously considered for the vacuum case do not contain a locally anti-de Sitter (AdS) spacetime \cite{daSilvaC}. Indeed, the locally AdS solution appears as a special branch disconnected from the general one \cite{Bicak}.  The advantage of our general static solution is that it is smoothly connected to a locally AdS spacetime, and in fact, this is achieved just doing $b=0$ in (\ref{Linet1}). Explicitly, we obtain
\begin{equation}\label{locAdS}
ds^2=dr^2+e^{2 r/l}\left(-dt^2+l^2 d\phi^2+dz^2\right),
\end{equation}
which becomes the background required for computing the conserved charges in subsection \ref{three}. 

\subsection{Local properties}
In order to obtain a deeper insight into the geometrical properties of the solution, we make use of an invariant characterization of the spacetimes. Spacetimes are usually classified according to the Petrov classification of their Weyl invariants.  Note that for analyzing the local properties it is enough to consider the static solutions because, as it will be shown in the next subsection, the stationary solutions can be obtained from a topological construction, and therefore they are locally equivalent. The general solution presented above, (\ref{Linet1}), is of type I (named normally algebraically general). However, as Linet pointed out in \cite{Linet}, a particular choice of the constants $K_0$, $K_1$ and $K_2$, makes the solution to be an algebraically special spacetime of type D. We find that, with the inclusion of the scalar field, i.e. by means of the constant $\a$, the Petrov type D spacetimes are no longer  determined only by those particular values of $K_i$, but by a range of values driven by $\a$. Namely, Petrov type D spacetimes are found for values of $K_i$ taken as any ordering of $\pm\frac {2} {3}\sqrt {4  - 3\alpha}$, $\mp\frac {1} {3}\sqrt {4 - 3\alpha}$, and $\mp\frac {1} {3}\sqrt {4 - 3\alpha}$, provided $0\leq\alpha < 4/3$. These type D spacetimes have a planar section (two $K_i$ are equal), which allows an additional symmetry. This fourth Killing vector corresponds to a rotation or a boost in this plane depending on its signature.  

A novel feature introduced by the scalar field, is a nontrivial Petrov type O subfamily. In fact, for $\alpha=\frac{4}{3}$, $b\neq 0$ and vanishing $K_i$, a conformally flat spacetime arises, and it is given by 
\begin{equation} \label{cf}
ds^2=dr^2+(e^{3 r/l}-b^2 e^{-3 r/l})^{2/3}(-dt^2+dz^2+l^2 d\phi^2).
\end{equation}
In other words, the scalar field gives rise to a wider family of spacetimes. This Petrov type O is a new subfamily parametrized by $b$, which strictly emerges due to the scalar field. In this case the number of isometries is enlarged to six since we are dealing with a conformally flat spacetime. It is remarkable to have such a number of symmetries in a space endowed with a matter source, in particular since for the vacuum (nontrivial)  case there are at most four Killing vectors \cite{daSilvaC}. For $b=0$ the scalar field is trivial ---it is a constant--- and (\ref{cf}) reduces to the locally AdS spacetime (\ref{locAdS}).

Studying the Weyl and Ricci scalars of the Newman-Penrose formalism it is shown that they are singular at the axis for the whole family of solutions, except in two cases.  The first one, corresponds to the CSI spacetimes, which will be discussed in Sec. \ref{csi}. The second case appears for a constant scalar field ($\alpha=0$) provided the constants $K_i$ take the values $\{\pm\frac{4}{3},\mp\frac{2}{3},\mp\frac{2}{3}\}$, or any permutation of them  \cite{Linet}. Since this special solution is regular at the axis, a change of the radial coordinate $r$ can be performed to prove that this type D solution is a black string. In fact, for $K_0=4/3$, and $K_1=K_2= -2/3$ the transformation reads 
\begin{equation} 
r= \frac{2 l}{3} \log\left[\frac{\rho^{3/2}+\sqrt{\rho^3-4 b l^3 }}{2 l^3} \right],
\end{equation}
yielding the black string
\begin{equation}
ds^2=-\left( \frac{\rho^2}{l^2}-\frac{4 l b}{\rho}\right)dt^2 + 
\frac{d\rho^2}{\displaystyle \frac{\rho^2}{l^2}-\frac{4 l b}{\rho}}+
\frac{\rho^2}{ l^{2}}dz^2+\rho^2 d\phi^2.
\end{equation}
Note that the original axis of symmetry at $r_0=l/3\log |b|$ is mapped to the horizon $\rho_+= 2^{2/3} l b^{1/3}$, and the new axis of symmetry is located at $\rho=0$. 
This black string was previously found by solving the Einstein field equations in \cite{Lemos}, and by using an adequate coordinate transformation in \cite{Bicak}.

\subsection{Topological construction of the rotating solution from a static one}
As explained in \cite{MacCallum}, a diagonal static metric with dependence on the spacelike coordinates $r$ and $z$,  and with the ``angular" coordinate stretched to infinity, can be locally equivalent but globally different to a stationary axisymmetric metric obtained from a topological identification in the static spacetime. This identification is defined by two essential parameters. This kind of essential parameters can not be removed by a permissible change of coordinates since they encode topological information. In this section we are going to build the stationary solution \eqref{ds2} with the metric coefficients \eqref{ds2comp}, using the procedure presented in \cite{MacCallum} in the particular case of cylindrical symmetry. 

Let us consider the static solution with scalar field
\begin{equation}\label{Linetscalar}
ds^2=-g_{0}(r)d\th^2+g_{1}(r) l^2 d\ph^2+g_{2}(r)dz^2+dr^2,
\end{equation}
where $g_{i}$ is given by \eqref{gs} in a coordinate system $(\th,r,z,\ph)$ with $\th\in(-\infty,\infty)$, $r\in[0,\infty)$, $z\in(-\infty,\infty)$ and $\ph\in(-\infty,\infty)$. Note that $\ph$ is not a compact coordinate. We perform a coordinate transformation on the $(\th,\ph)$ plane given by
\begin{equation}\label{ct}
\th=\b_0 \phi+\b_1 t,\qquad
\ph=\a_0 \phi+\a_1t,
\end{equation}
where $\a_0$, $\a_1$, $\b_0$ and $\b_1$ are parameters. This transforms \eqref{Linetscalar} into  \eqref{ds2comp} by defining these parameters as follows
\begin{eqnarray}\label{parameters}
\a_0&= \sqrt{a_0}, \quad \a_1 = \displaystyle -\frac{\sqrt{a_1}}{l},\nonumber \\
\b_0&= -\sqrt{b_0}, \quad
\b_1 =\displaystyle\frac{\sqrt{b_1}}{l}.
\end{eqnarray}
As shown in \cite{MacCallum}, $\alpha_1$ and $\beta_1$ are not essential parameters, and they can be set as $\alpha_1=0$ and $\beta_1=1$. On the contrary, $\alpha_0$ and $\beta_0$ are essential. However, after a topological identification, which transforms the $(\th,\ph)$ plane  into a cylinder, one can fix the period of the angular coordinate $\phi$ to $2\pi$ by choosing $\alpha_0= 1$.  Since that in (\ref{Linetscalar}) all the coordinates are not compact, $b$ can be absorbed by rescaling the coordinates. After identification, $\ph$ becomes periodic and $b$ has a topological meaning. The parameter $\alpha_0$ plays the same topological role, and in fact it redefines $b$. Therefore, without loss of generality $\alpha_0$ can be fixed, but not simultaneously with $b$.   In other words, since from the beginning the static solution contains an arbitrary conicity parameter $b$, the constant $\alpha_0$ can be fixed.  Going back to relations \eqref{parameters} we find that $a_0=1, a_1=0$ and $b_1=l^2$ reproduce the set of values chosen for $\alpha_0, \alpha_1$ and $\beta_1$. Then, after fixing the period as $2 \pi$  there is just one essential parameter $\beta_0$ in the transformation, which will be named $-a$ hereafter. Then,  the transformation (\ref{ct}) reduces to
\begin{equation}\label{trans}
\th=t-a \phi,\qquad \ph=  \phi.
\end{equation}
 In summary, a topological construction can bring the solution \eqref{Linetscalar} into a locally equivalent, but globally different, solution by doing the transformation (\ref{trans})
to get
\begin{equation}\label{Linetsol}
ds^2=-g_0(r)(dt-a  d\phi)^2+g_1(r) l^2 d\phi^2+g_2(r)dz^2+dr^2.
\end{equation}
Transformation \eqref{trans} is not a proper coordinate transformation, since it converts an exact 1-form into a closed but not exact 1-form, as was discussed in detail in \cite{Stachel}. Hence, (\ref{trans}) only preserves the local geometry, but not the global one. Therefore, the resulting manifold is globally stationary but locally static. Hereafter, we will consider \eqref{Linetsol} instead of \eqref{ds2comp} as the general solution, because it already contains all the local and global essential information.

\subsection{Asymptotic behavior}

In order to display the asymptotic behavior of the fields, it is convenient to use the coordinate $\rho= l e^{r/l}$. In this way, the behavior at large $\rho$ is given by 
\begin{equation}\label{asymp}
\begin{split}
g_{tt}(\rho)&=-\frac{\rho^2}{l^2} + \frac{2 b l K_0}{\rho}+O(\rho^{-4}),\\
g_{\p\p}(\rho)&=\rho^2 (1 - \frac{a^2}{l^2})+\frac{ 2 l b (-l^2 K_1+a^2 K_0)}{\rho}+O(\rho^{-4}),\\
g_{t\p}(\rho)&=\frac{\rho^2 a}{l^2}-\frac{ 2 b l a K_0}{\rho}+O(\rho^{-4}),\\
g_{zz}(\rho)&=\frac{\rho^2}{l^2}-\frac{ 2 b l K_2}{\rho}+O(\rho^{-4}), \qquad  g_{\rho \rho}(\rho)=\frac{l^2}{\rho^2},\\
\Phi(\rho)&= \Phi_0+ \sqrt{\frac{2 \alpha}{\kappa}}\frac{ b l^3}{\rho^3}+ O(\rho^{-9}).
\end{split}
\end{equation}
One can note that the metric asymptotically approaches a locally AdS spacetime, as the scalar field becomes constant.  The background is fixed by setting $a=b=\alpha= \Phi_0=0$, which corresponds to a locally AdS spacetime.

\subsection{Mass and angular momentum}\label{three}
The mass and angular momentum of the solutions are determined using the Regge-Teitelboim method \cite{ReggeTeitelboim}. In the canonical formalism, the generator of an asymptotic symmetry associated to the vector $\xi=(\xi^{\perp},\xi^{i})$ is built as a linear combination of the constraints $\mathcal{H}_{\perp}, \mathcal{H}_{i}$, with an additional surface term $Q[\xi]$ 
\begin{equation}
H[\xi]=\int d^{3} x \left(  \xi^{\perp} \mathcal{H}_{\perp}+\xi^{i}
\mathcal{H}_{i}\right)  +Q[\xi].
\end{equation}
A suitable choice of this surface term attains the generator has well-defined functional derivatives with respect to the canonical variables \cite{ReggeTeitelboim}. The surface term $Q[\xi]$ is the conserved charge under deformations $\xi$ provided the constraints vanish. For the action (\ref{action}), the variation of $Q[\xi]$ is given by
\begin{align}\label{de}
&\d Q[\xi]=\oint d^{2}S_{l}\left[ \frac{G^{ijkl}}{2 \kappa}(\xi^{\bot}\d g_{ij;k}-{\xi^{\bot}}_{,k} \d g_{ij})+2 \xi_{k}\d \pi^{kl}\right. \nonumber\\&
\! \! \!\left.+(2 \xi^{k}\pi^{jl}\!-\!\xi^{l}\pi^{jk})
\d g_{jk}\!-\! (\sqrt{g} \xi^{\bot}g^{lj}\Phi_{,j}\!+\!\xi^{l}\pi_{\Phi})\d\Phi \right],
\end{align}
where $G^{ijkl}\equiv\sqrt{g}(g^{ik}g^{jl}+g^{il}g^{jk}-2g^{ij}g^{kl})/2$. The canonical variables are the spatial metric $g_{ij}$ and the scalar field $\Phi$ together with their respective conjugate momenta $\pi^{ij}$ and $\pi_\Phi$.

To evaluate $\d Q[\xi]$ we consider as asymptotic conditions just the asymptotic behavior of the solutions with a negative cosmological constant (\ref{asymp}), where the integration constants $K_{i}, a, b,  \alpha$ are allowed to be varied. The additive constant of the scalar field $\Phi_0$  is  considered as a fixed constant without variation, in order to save the asymptotic scale invariance\footnote{For $\delta\Phi_0 \neq 0$,  $\d Q[\xi]$ contains a term proportional to $\oint d^{2}S  \xi^t\sqrt{\a} b \d \Phi_0$. The integration of this term requires a boundary condition relating $\Phi_0$ with  $\a$ and $b$.}. Since the solution is in the comoving frame along $z$, the corresponding momentum $Q[\partial_z]$ vanishes. Then, the only nonvanishing charges are those associated to symmetry under time translations and the rotational invariance, the mass and angular momentum, respectively. Defining $ q[\xi]$ as the charge by unit length $ Q[\xi]=\int  q[\xi] dz$, we can obtain from (\ref{asymp}) and (\ref{de}),  the explicit form of $\d q[\xi]$ 
\begin{align}
\d q[\xi]&= \frac{6\pi }{\kappa}\left[ -\xi^t \delta (b(K_1+K_2))+\xi^\phi \delta (a b(K_1-K_0)) \right].
\end{align}
Thus,  using $\kappa= 8 \pi G$, the mass $M=q[\partial_t]$ and angular momentum $J=q[\partial_\phi]$ per unit length
are 
\begin{align}
M&=\frac{3 b}{4 G}K_0, &J&=\frac{3 a b}{4 G }(K_1-K_0).
\end{align}
These global charges are defined up to an additive constant without variation. 
In order to set the locally AdS spacetime (\ref{locAdS}) as a background, these additive constants must be chosen to be null.

As we can see from the expression for the angular momentum, there are two manners of turning off the angular momentum. The first one is by doing $a=0$, which cancels the off-diagonal term $g_{t\p}$ in the metric. The second way is less obvious, since it is achieved by considering $K_0=K_1$. Indeed, this  particular choice of the parameters yields a static solution of type D. This can be shown from the coordinate transformation 
\begin{equation}
d\phi \rightarrow d\phi+\frac{a}{(a^2-l^2)}d t, \qquad d t \rightarrow d t.
\end{equation}
As analyzed  in \cite{MacCallum}, this transformation contains an inessential parameter $\a_1= a/(a^2-l^2)$, which does not change the topology. Therefore, the solution with $K_0=K_1$ is no just locally equivalent to the static solution, but also globally.

\section{Stationary cylindrically symmetric solutions with $\Lambda=0$}\label{four}
In this section the field equations are integrated using the same procedure as in \cite{Linet}, but assuming from scratch a vanishing cosmological constant. We show that the limit $\Lambda \rightarrow 0$, or  equivalently $l\rightarrow\infty$,  in the configurations of Sec. \ref{two}, does not provide all the solutions coming from a direct integration of the field equations. Two classes of solutions are obtained. The first type corresponds to solutions that match the limit $\Lambda \rightarrow 0$ in the configurations introduced in Sec. \ref{two}, and they are dubbed as Levi-Civita type spacetimes. The second type is formed by spacetimes having all their invariants constant. These two types will be analyzed in detail below. The discussion in this section is focused on static solutions. The topological construction explained in Sec. \ref{two} does not depend on the value of the cosmological constant, and in consequence, the stationary solutions for $\Lambda=0$ can be obtained from the improper transformation (\ref{trans}). Since (\ref{trans}) is a local transformation, the static configuration and its stationary counterpart share the same local properties.

\subsection{Levi-Civita type spacetimes} \label{LCs}
In this subsection, we show a Levi-Civita type spacetime in presence of a massless scalar field. We consider the most general static cylindrical metric
\begin{equation}\label{Linetnull}
ds^2=-g_{0}(r)dt^2+g_{1}(r) d\phi^2+g_{2}(r)dz^2+dr^2,
\end{equation}
with a scalar field depending just on the radial coordinate $r$.
 Following \cite{Linet}, the Einstein field equations can be put in a very simple form
\begin{eqnarray}
\left(\left(\frac{u}{g_i}\right)g_i^{\prime}\right)'&=&0,\quad i=0,1,2,\label{eqh}\\
\frac{g_0^{\prime}g_1^{\prime}}{g_0g_1}+\frac{g_1^{\prime}g_2^{\prime}}{g_1g_2}+\frac{g_2^{\prime}g_3^{\prime}}{g_2g_3}&=&2\kappa\Phi'^2,\label{eqnh}
\end{eqnarray}
with $u\equiv \sqrt{g_0 g_1 g_2}$. Using the sum of all the equations in \eqref{eqh}, and the definition of $u$, one obtains $u''=0$, so that
\begin{equation} \label{usol}
u(r)=K r+u_0,
\end{equation}
where $K$ and $u_0$ are integration constants. Substituting  (\ref{usol}) in \eqref{eqh}, and choosing the axis at $r=0$, we obtain
\begin{equation}\label{gsss}
g_i(r)=r^{2/3+K_i}g_i^0,
\end{equation}
where $K_i$ and $g_i^0$ are integration constants fulfilling  $g_0^0g_1^0g_2^0=K^2$. Moreover, replacing (\ref{gsss}) in the definition of $u$ we obtain the same relation for the $K_i$ given by Eq. (\ref{relnorm}). The scalar field is obtained from (\ref{eqnh}), $\Phi=\sqrt{\a/2\k}\log(r/r_0)$, where $\alpha$ is related with $K_i$ in the same way as in \eqref{relesp}, and $r_0$ is an arbitrary constant.  
The algebraic relations (\ref{relnorm}) and \eqref{relesp} determine two essential constants related to the gravitational and scalar field strengths.  Since $\phi$ is an angular coordinate with a given period, the constant $g_1^0$ cannot be absorbed by a rescaling of this coordinate keeping the same period. Then, $g_1^0$ is a third essential parameter, and plays a topological role in the same way as $b$ in the previous section. The transformation (\ref{trans}) provides the fourth essential parameter for the stationary solution.

Since $K_i$ and $\alpha$ must satisfy (\ref{relnorm}) and \eqref{relesp}, they are bounded in the same way as was established in Section \ref{two}.
 
As in Sec. \ref{two}, we study the local properties through the Petrov classification. Normally the solution is algebraically general as occurs in vacuum \cite{Griffiths}, but algebraically special spacetimes are also possible to be found.  The scalar field parametrizes three families of type D spacetimes, which will be described in Table 1. Two of these families ($S_1$ and $S_2$) are allowed only for a nonvanishing scalar field, while the third one ($S_3$) reduces to the three known vacuum type D Levi-Civita spacetimes by switching off the scalar field and by circular permutations of $K_i$. A nontrivial type O spacetime emerges strictly from the scalar field. In this case $K_0=K_1=K_2=0$ and $\alpha=4/3$ yielding the conformally flat metric
\begin{equation}
ds^2=dr^2+r^{2/3}(-dt^2+dz^2+g_1^0 d\phi^2).
\end{equation}
This is the counterpart with $\Lambda=0$ of the conformally flat spacetime described in (\ref{cf}). 

\begin{table}[h]
\begin{tabular}{ccccc}
\hline \hline
 & $K_0$ & $K_1$ & $K_2$ & $\alpha$ \\ 
\hline 
$S_1$ & $\frac {2} {3}\sqrt {4  - 3\alpha}$ & $-\frac {1} {3}\sqrt {4  - 3\alpha}$ & $-\frac {1} {3}\sqrt {4  - 3\alpha}$ & $(0, \frac {4}{3})$ \\ 
$S_2$ & $-\frac {2} {3}$ & $\frac {1} {3}\pm\sqrt {1  - \alpha}$ & $\frac {1} {3}\mp\sqrt {1  - \alpha}$ & $(0,1]$ \\ 
$S_3$ & $-\frac {2} {3}\sqrt {4  - 3\alpha}$ & $\frac {1} {3}\sqrt {4  - 3\alpha}$ & $\frac {1} {3}\sqrt {4  - 3\alpha}$ & $[0, \frac {4}{3})$ \\  
\hline \hline
\end{tabular} 
\caption{Petrov D spacetimes for $\Lambda=0$. The constants $K_i$ are classified in three sets, and depend on the amplitude of the scalar field $\alpha$.  Within each set $K_0$, $K_1$ and $K_2$ can be taken in any order. The last column shows the range of $\alpha$ allowed for each set. The first two sets are exclusive for a non-constant scalar field ($\alpha \neq 0$), and the third one also includes a trivial scalar field.}
\end{table}

It is found that the nonvanishing  components of the Riemann tensor ${R^{\mu \nu}}_{\lambda \rho}$ and Kretschmann scalar are proportional to $r^{-2}$ and $r^{-4}$, respectively. Then, the spacetime is asymptotically locally flat. 

Until now, we have assumed a nonvanishing $K$. However, when we consider $K=0$, i.e $u=u_0$, equations \eqref{eqh}  drastically modify the functional form of $g_i(r)$. This new branch of solutions, which is not provided by the limit $\Lambda \rightarrow 0$ in Sec. \ref{two}, are analyzed in next subsection.
\subsection{CSI spacetimes}\label{csi}
In general, the Levi-Civita type spacetimes discussed above possess curvature invariants which are singular at $r=0$. However, it is possible to find regular spacetimes, i.e spacetimes free of any curvature singularity, where in addition, all polynomial scalar invariants constructed from the Riemann tensor and its covariant derivatives are constant. These spacetimes are known as constant scalar invariant (CSI) spacetimes. In this subsection, a non-trivial CSI spacetime due to the presence of the scalar field is presented. It is found that it is required to switch off the cosmological constant in order to get this class of spacetimes. This case is of particular interest since it provides a  non-vacuum solution with constant curvature scalars. For simplicity, only the static cases will be considered, since the stationary CSI spacetimes can be obtained by performing the coordinate transformation \eqref{trans}.

This special spacetime comes from considering $K=0$  in \eqref{usol}. As mentioned before, this branch  of solutions is not smoothly connected with that shown in subsection \ref{LCs}. In fact, the solutions of the equations \eqref{eqh} are given by exponentials, 
\begin{equation} \label{gcsi}
g_i(r)=g_i^0e^{K_i r},
\end{equation}
and the scalar field, obtained from (\ref{eqnh}), is a linear function, $\Phi(r)=\Phi_0+\sqrt{\alpha /(2 \kappa)}r$. The integration constants satisfy the algebraic relations,
\begin{equation}\label{CSIrel}
\begin{split}
K_0+K_1+K_2&=0,\\
K_0 K_1+K_1 K_2+K_2 K_0&=\alpha.
\end{split}
\end{equation}
Thus, from \eqref{CSIrel} we obtain,
\begin{eqnarray}
K_0&=&\frac{1}{2}(-K_2\pm\sqrt{-3 (K_2)^2-4\a}),\\
K_1&=&\frac{1}{2}(-K_2\mp\sqrt{-3 (K_2)^2-4\a}).
\end{eqnarray}\\[-7pt]
Note that the reality condition of the line element demands $\a<-\frac{3}{4}(K_2)^2$ and as a consequence the scalar field becomes imaginary. This means that the presence of a phantom scalar field makes possible to remove curvature singularities present in the vacuum solutions. The Petrov classification indicates that these spacetimes are type D.

In order to verify that these spacetimes are indeed CSI spacetimes, we make use of a theorem proved in \cite{Coley1,Coley2}. The theorem states that any four dimensional locally homogeneous spacetime is a CSI spacetime. The line element \eqref{Linetnull} with the metric coefficients (\ref{gcsi}) has three trivial Killing vectors $\partial_t$, $\partial_z,$ and $\partial_\phi$. However, it is possible to find a fourth Killing vector given by
\begin{equation}\label{tkv1}
\xi^{(4)}=(-\frac{1}{2}K_0t,-\frac{1}{2}K_1\phi,-\frac{1}{2}K_2 z,1),
\end{equation}
in the coordinate system $(t,\phi,z,r)$, which in addition to the trivial ones, form a transitive group of isometries. Therefore, this spacetime is locally homogeneous. 

\section{Concluding remarks}
In this paper, the general stationary cylindrically symmetric solution of Einstein-massless scalar field system with a non-positive cosmological constant has been found, and its local and global properties has been studied.  Four integration constants are essential parameters for the general solution. This means that these parameters encode all the relevant physical information. One is the amplitude of the scalar field, which beside a second one present in the metric, characterize the gravitational field strength. The other two parameters have a topological origin, since they appearing  in the improper gauge transformation that allow us to obtain the stationary solution from the static one. The meaning of these parameters can be also analyzed from the expressions for the mass and  angular momentum of the solutions with a negative cosmological constant. 

The Petrov classification was performed to explore the effects of the scalar field on the vacuum solutions for a negative and a vanishing cosmological constant. The inclusion of the scalar field enlarges the family of solutions in comparison with the vacuum case. Thus, type D solutions are now parametrized by the amplitude of the scalar field and nontrivial type O solutions have been found in presence of nonvanishing scalar field. These conformally flat solutions endowed with a matter field have six Killing vectors. Note that in the vacuum case, there are not type O solutions apart from the trivial ones, the locally Minkowski (for $\Lambda=0$) and the locally AdS spacetime (for $\Lambda<0$).

Other interesting case occurs for $\Lambda=0$. There are  special type D solutions which are possible only if the scalar field is present. We have shown that these spacetimes have a fourth Killing vector, which completes a transitive group of isometries, and consequently they are locally homogeneous.  Thus, these solutions become CSI spacetimes dressed by a phantom scalar field.

\acknowledgments

We thank Eloy Ay\'on-Beato, Patricia Ritter, Marco Astorino and Ricardo Troncoso for helpful discussions. C. E. thanks CONICYT for financial support. This work has been partially funded by the  Fondecyt grants 1121031 and 1130658.
The Centro de Estudios Cient\'{\i}ficos (CECs) is funded by the Chilean Government
through the Centers of Excellence Base Financing Program of Conicyt.

\end{document}